\documentstyle[12pt]{article}
\begin{document}
\renewcommand{\thesection}{\Roman{section}}
\bibliographystyle{unsrt}
\newcommand{\nwc}{\newcommand}
\nwc{\be}{\begin{equation}}
\nwc{\ee}{\end{equation}}
\nwc{\bea}{\begin{eqnarray}}
\nwc{\eea}{\end{eqnarray}}
\nwc{\ba}{\begin{array}}
\nwc{\ea}{\end{array}}
\nwc{\rtr}{\rangle}
\nwc{\ltr}{\langle}
\nwc{\ket}[1]{|#1\rtr}
\nwc{\bra}[1]{\ltr#1|}
\nwc{\scal}[2]{\bra{#1}#2\rtr}
\nwc{\dagg}{\mbox\footnotesize{\dag}}
\nwc{\emp}{\emphasize}
\nwc{\lb}{\label}
\nwc{\rf}[1]{~(\ref{#1})}
\nwc{\ci}[1]{~\cite{#1}}
\nwc{\pa}{\partial}
\nwc{\paf}[2]{\frac{\pa#1}{\pa#2}}
\nwc{\ra}{\rightarrow}
\nwc{\Tr}{\mbox{\rm Tr}}
\nwc{\real}{\mbox{\rm Re}}
\nwc{\im}{\mbox{\rm Im}}
\nwc{\bino}[2]{\mbox{$\left(\begin{array}{c}#1\\#2\end{array}\right)$}}
\def\a{\alpha}
\def\b{\beta}
\def\e{\epsilon}
\def\l{\lambda}
\def\m{\mu}
\def\f{\phi}
\def\p{\pi}
\def\t{\theta}
\def\D{\Delta}
\def\O{\Omega}
\def\s{\sigma}
\nwc{\eps}{\epsilon}
\nwc{\br}{\mbox{\bf{R}}}
\nwc{\bc}{\mbox{\bf{C}}}
\nwc{\cz}{{\cal Z}}
\nwc{\cd}{{\cal D}}
\title{Path integrals for spinning particles, stationary phase and the
Duistermaat-Heckman theorem.}
\author{E. Ercolessi$^*$, G. Morandi$^\dagger$, F. Napoli$^\ddagger$
and P. Pieri$^\dagger$\\
{\it\small $^*$ Dipartimento di Fisica and INFM, Universit\`a di
Bologna,}\\
{\it\small 46 Via Irnerio, I-40126, Bologna, Italy.}\\
{\small\it$^\dagger$ Dipartimento di Fisica, INFM and INFN,
Universit\`a di Bologna,}\\
{\it\small 46 Via Irnerio, I-40126, Bologna, Italy.}\\
{\it\small $\ddagger$ Istituto di Fisica di Ingegneria and INFM,
Universit\`a di Genova,}\\
{\it\small I-16100 Genova,Italy.}}
\date{}
\maketitle
\begin{abstract}
We examine the problem of the evaluation of both the propagator and of
the partition function of a spinning particle in an external field at
the classical as well as the quantum level, in connection with the
asserted exactness of the saddle point approximation (SPA) for this
problem. At the classical level we argue that exactness of the SPA stems
from the fact that the  dynamics (on the two--sphere $S^2$) of a
classical spinning particle in a magnetic field is the reduction from
$\br^4$ to $S^2$ of a linear dynamical system on $\br^4$. At the quantum
level, however, and within the path integral approach, the restriction,
inherent to the use of the SPA, to regular paths clashes with the fact that no
regulators are present in the action that enters the path integral.
This is shown to lead to a prefactor for the path integral that is
strictly divergent except in the classical limit. A critical comparison
is made with the various approaches to the same problem that have been
presented in the literature. The validity of a formula given in
literature for the spin propagator is extended to the case of motion in an
arbitrary magnetic field.
\end{abstract}
\section{Introduction.}

   Since the early days of path integration, how to do a path
integral for spinning particles was recognized\ci{fey} as one of the major
difficulties of the formalism. Schulman\ci{schu68}(but see also\ci{schu})
made a first attempt towards a  formulation of a path integral for spinning
particles, one which was however rather a related path integral, namely
that for a spinning top.

   Much progress has been made since with the systematic use, initiated
by Klauder\ci{kla78,kla79}, of the resolution of the identity associated with
spin--coherent states\ci{kla,per} in the discretized-time (or time-sliced)
approach to the path integral. Path integral quantization using coherent
states will be discussed extensively in the sequel. Other versions of the
path integral not making an explicit use of coherent states have been
discussed instead in\ci{nie88,sto89} (see also\ci{joh89}).

   Already at the classical level, a spin (classically a vector ${\vec S}$
of fixed magnitude $s$) presents some peculiarities as a dynamical system.
While the Hamiltonian description is essentially straightforward if one
assumes the Poisson brackets:
\be
\{S_i,S_j\}=\eps_{ijk}S_k
\lb{pb}
\ee
among the components of the spin vector, it has been pointed out by
Balachandran {\em et al}.\ci{bal} that no global Lagrangian description can be
given as long as one sticks to the natural configuration space of the spin,
which is the compact two-sphere $S^2$. The same authors have shown that a
global Lagrangian can be associated with a classical spin by lifting the
description of the system from $S^2$ to the group manifold of SU(2), which
is the three-sphere $S^3$, as follows. Considering the usual spin-1/2
representation of SU(2), one can define a vector ${\vec S}$ in $S^2$ via
the Hopf map, i.e.:
\be
{\widetilde S}={\vec S}\cdot{\vec \sigma}=s g\sigma_3g^{-1}\; ,
\lb{hp}
\ee
with ${\vec \sigma}=(\sigma_1,\sigma_2,\sigma_3)$ the Pauli matrices. Then
it can be shown\ci{bal} that the (global) Lagrangian on TSU(2):
\be
L= i s \Tr ( \sigma_3 g^{-1}{\dot g}) - \frac{\mu}{2}\Tr({\widetilde S}
{\widetilde B}) \; ,
\lb{bl}
\ee
where $\mu$ is the Bohr magneton and
${\widetilde B}={\vec B}\cdot{\vec \sigma}$,
 yields the correct equations of motion for a classical spin
in an external magnetic field ${\vec B}$. The same equations can be derived
of course at the Hamiltonian level using the Poisson brackets\rf{pb} and the
Hamiltonian:
\be
   H=\mu {\vec S}\cdot{\vec B} \label{sh} \; .
\ee

As it is clear from the fact that the spin is recovered via the Hopf
projection\rf{hp}, this approach introduces an extra, nondynamical U(1)
gauge degree of freedom. Indeed, under: $g\ra g \exp[i\gamma \sigma_3/2]$,
\rf{hp} is invariant while $L$ changes by a total time derivative (i.e. it
is ``weakly invariant"\ci{bal}):
\be
L\ra L-s{\dot \gamma} \; .
\ee
This is also evident if we parametrize SU(2) with the Euler angles as
\be
g=e^{-i\phi\sigma_3/2}e^{-i\theta\sigma_2/2}e^{-i\gamma\sigma_3/2}
\ee
($0\leq \theta\leq \pi, 0\leq \phi\leq 2\pi,0\leq \gamma\leq 4\pi$),
which yields the (local)
Lagrangian:
\be
L=s( {\dot \phi}\cos\theta + {\dot\gamma}) +\mu {\vec S}\cdot{\vec B}
\; . \lb{lc}
\ee

    Although it can be shown\ci{biagio} that this extra gauge degree of
freedom has interesting consequences for the path integral
quantization of\rf{bl}, and namely that it leads in a straightforward
way to spin quantization (i.e.: $2s/\hbar$=integer), we would like to
stress here the fact that\rf{bl} (or\rf{lc} for that matter) being linear in
the time derivatives, a spinning particle is described as a constrained
dynamical system. Canonical quantization requires then the use of Dirac's
theory\ci{dir} of constraints, and it has been shown by Balachandran
{\em et al}.\ci{bal} that this does indeed yield the correct quantum mechanical
description of the spin including spin quantization. As pointed out, e.g.,
in Ref.\ci{sto89}, it turns out that the (semiclassical) Bohr-Sommerfeld
 quantization
is exact for a spin Hamiltonian. This is of course a strong indication that
the saddle point, or stationary phase,  approximation (henceforth
referred to as SPA) to the path integral for, say, the propagator
should be exact as well.

Concerning  the path integral approach and the legitimacy of evaluating the
path integral within the SPA, which is the main problem which
the present paper addresses to, a Lagrangian of the form\rf{lc} poses
another problem, and namely that, the classical equations of motion being
first-order in time, the saddle point problem becomes {\em overdetermined},
as the solutions of the saddle point equations have to obey two boundary
conditions and not a single one.
Different ways to overcome this problem have been proposed in the literature,
and notably by Klauder\ci{kla78,kla79}, Keski-Vakkuri {\em et al}.\ci{vak91}
and
Suzuki\ci{suz83}.

 This paper is devoted mainly to the discussion of two questions, namely to
 that of the legitimacy of some approximations that are currently
 made in the coherent state formulation of the path integral for spins and
 of the apparently surprising fact, that has also been widely discussed in
 the literature, that the SPA to the path integral yields the exact
result for both the
 propagator and/or the partition function of a quantum spin. While the SPA
 is almost trivially exact for quadratic Hamiltonians, here the apparently
 surprising fact is that it turns out to be exact also for the Hamiltonian
 (\ref{sh}) and/or the Lagrangian (\ref{lc}) that are far from being quadratic.
Exactness of the SPA implies of course that we discuss the applicability
to the present problem of the Duistermaat-Heckman (DH) theorem which provides
a rigorous framework for the discussion of the exactness of the SPA.

 The paper is organized as follows. In the Appendix we briefly review
the DH theorem and the SPA. In Section II
we apply the latter to the calculation of the partition function of a
classical spin and show that the deep reason of the validity of such an
approximation is that a classical spin,
 when viewed as a dynamical system, can be shown to result from the reduction
of a linear dynamical system from $\br^4$ to $S^2$. In Section III we begin by
discussing briefly the use of spin--coherent state path integral
approach to the calculation of the propagator and/or of the partition
function of a quantum spin.
We show there that the approximation currently used in literature
restricting the paths in the functional
integral to be continuous leads to incorrect and diverging prefactors.
Then we review briefly the approaches of Klauder and
Suzuki to the same problem. We clarify the origin of some apparently
mysterious terms that are added to the action in Klauder's approach. Then
we analyze why the SPA yields the exact result
in Suzuki's approach, by arguing that his main result stems simply from
the exactness of the Ehrenfest theorem in the present context.
In Section IV we discuss in more detail the coherent state approach by
using the holomorphic representation for the latter,
applying the complex SPA to the discrete version of the spin path
integral. The final Section V is devoted to a general
discussion and to some conclusions.

\section{The classical spin.}

Let us start our considerations from a {\it classical} (nonrelativistic)
spin in a time-independent magnetic field $\vec{B} = (B_1,B_2,B_3)$.

Since the only degree of freedom for a spin is its direction, we describe it
by means of a three dimensional vector $(S_1,S_2,S_3)\in \br^3$
of fixed norm: $S_1^2+S_2^2+S_3^2=s^2$, where the three classical variables
$S_j$ ($j=1,2,3$) satisfy the Poisson brackets (\ref{pb}).
Thus the phase space for a classical spin is the two-dimensional manifold
$S^2(s)$, equipped with the symplectic two-form
\be
\O = \frac{1}{2s^2}
\e_{ijk} S_i dS_j \wedge dS_k \; .
\ee

The Hamiltonian that describes a classical spin in a magnetic field is
given by (\ref{sh}). For simplicity we will
set $\m\equiv 1$ in the following so that
\be
H= \vec{B} \cdot \vec{S} \label{ham} \; .
\ee

The Hamiltonian vector field $\D$ associated to (\ref{ham}), and determined by
the symplectic form $\O$, is given by
\be
\D = \e_{ijk} B_j S_k \frac{\partial}{\partial S_i} \; ,
\ee
so that the classical equations of motion, $i_\Delta\Omega=dH$, read
\be
\dot{S}_i = \e_{ijk} B_j S_k \label{em} \; .
\ee
Without any loss of generality, we will set $\vec{B} = (0,0,B)$
($B>0$) from now on. In this case,
the Hamiltonian (\ref{ham}) and the equations of motion (\ref{em}) assume a
very simple form if we use spherical coordinates $S_1= s \sin\t \cos\f$,
$S_2=s \sin\t \sin\f$, $S_3= s\cos\t$. They become respectively
\be
H= s B \cos\t  \label{hs}
\ee
and
\bea
\cos\t \dot{\t} =0 \label{ems}\\
\sin\t (\dot{\f} - B ) = 0 \; . \nonumber
\eea
The latter can be easily integrated and one sees that the classical orbits
are circles parallel to the equator, the spin precessing about the
magnetic field with a period $\tau = \frac{2\p}{B}$.

In spherical coordinates, it is also very easy to compute the {\it exact}
partition function for the Hamiltonian (\ref{ham}):
\be
{\cal Z}_{cl}(\b) = \int_{S^2(s)} \O e^{-\b H} \; . \label{pf}
\ee
Indeed one gets:
\bea
{\cal Z}_{cl}(\b) &=& s \int_0^{2\p} d\f \int_0^{\p} \sin\t d\t
e^{-\b s B \cos\t} \label{pfs}\\
&=& \frac{2\p}{\b B} \left( e^{\b s B} - e^{-\b s B} \right)
\; . \nonumber
\eea
{}From (\ref{pfs}), one recognizes that the classical partition function
can be written as the weighted sum of two terms, each given by the
evaluation of $e^{-\b H}$ at the two critical points
$\t=\p$ and $\t=0$ of the Hamiltonian for which
$\D_{\vec{S}=(0,0,\pm s)}=0$.
In addition, it is not hard to check that the weights ($\pm
\frac{2\p}{\b B}$) are exactly the ones coming from the calculation of
the contributions to the integral (\ref{pfs}) of the gaussian fluctuations
around the stationary points of $H$.

Everybody is familiar with such a result whenever dealing with
(multidimensional) harmonic oscillators or in general with a quadratic
Hamiltonian on the linear manifold $\br^{2n}$. Even if the Hamiltonian
(\ref{ham}) in not of this kind, the SPA is exact as well.
A spin in a magnetic field is in fact the simplest (nontrivial)
application of the Duistermaat-Heckman theorem\ci{dh,bv},
which establishes under which conditions a phase-space integral,
such as (\ref{pf}), can be evaluated exactly in the SPA.
We refer to the Appendix for a review of the
Duistermaat-Heckman theorem and for the proof of its applicability to the
system of a spin in a magnetic field. Here we recall only that this
result holds essentially for the two following geometrical reasons:\\
1) the Hamiltonian (\ref{hs}) is invariant under an $U(1)$-action, given
by rotations about the third-axis (i.e. the axis of the constant magnetic
field);\\
2) the associated Hamiltonian vector field, given by
\be
\D= B \left(- S_2 \frac{\partial}{\partial S_1} +
S_1 \frac{\partial}{\partial S_2} \right)\label{16}
\ee
is proportional to the generator of this $U(1)$-action. This is clear in
spherical coordinates, where $\Delta=B\frac{\partial}{\partial \f}$.
However while\rf{16} defines it globally on $S^2(s)$, the sperical
coordinate representation becomes singular at $\t=0,\,\pi$.\\

The Duistermaat-Heckman theorem gives some abstract mathematical conditions
for the SPA to be exact. In the following we
will show that for a spin in a
magnetic field there is however a deeper reason why this holds: the dynamical
system that describes a spin in magnetic field is the reduction of a bigger
system which is described by a quadratic Hamiltonian  on the linear
manifold $\br^4$\ci{mssv}. Such a situation has already been
considered
in the literature (see\ci{glmv} and references therein), even if not
much in the context of the SPA, but mainly in the
context of integrable systems, for which there exists the conjecture\ci{glmv}
that every integrable system is the reduction of a bigger
{\it linear} dynamical system.
To show what happens in the case of a spin,
let us consider the linear manifold
$\br^4=\{(x_1,x_2,x_3,x_4)\}= \bc^2 = \{(z_1,z_2): z_1=
x_1+i x_2,z_2=x_3+ix_4 \}$, equipped with the symplectic two-form
$\bar{\O} = \frac{1}{2s} (dx_1 \wedge dx_2 + dx_3 \wedge dx_4) =
\frac{1}{2s} \frac{1}{2i} (dz_1^*\wedge dz_1 + dz_2^* \wedge dz_2)$.

There is a natural action of SU(2) on $\br^4=\bc^2$, which is
simply given by left multiplication:
\begin{displaymath}
\left[
\begin{array}{cc}
\a & -\b^* \\ \b & \a^* \end{array} \right]
\left[
\begin{array}{c}
z_1 \\ z_2 \end{array} \right]
= \left[
\begin{array}{c}
\a z_1 -\b^* z_2 \\ \b z_1 + \a^* z_2 \end{array} \right] \; ,
\end{displaymath}
where $\left[ \begin{array}{cc} \a & -\b^* \\ \b & \a^*
\end{array} \right]$, with $|\a|^2 + |\b|^2 =1$, is an element of SU(2)
in the fundamental representation.
This action is symplectic and Hamiltonian\ci{amr}, and its Lie
algebra is spanned by the vector fields:
\bea
\bar{\chi}_1 =  x_4 \partial_1 - x_3 \partial_2 +
x_2 \partial_3 - x_1 \partial_4  \label{vf} \\
\bar{\chi}_2 =  - x_3 \partial_1 - x_4 \partial_2 +
x_1 \partial_3 + x_2 \partial_4 \nonumber \\
\bar{\chi}_3 =  x_2 \partial_1 - x_1 \partial_2 -
 x_4 \partial_3 + x_3 \partial_4  \; . \nonumber
\eea
Hence it is easy to prove that the linear vector field $\bar{\D}=
\sum_{j=1}^3 B_j \bar{\chi}_j$ is a Hamiltonian vector field with a
quadratic Hamiltonian, namely:
\be
\bar{H} = \sum_{j=1}^3 B_j f_j \; , \label{hr}
\ee
where
\bea
f_1 = \frac{1}{2s} ( x_1 x_3 + x_2 x_4 ) = \frac{1}{2s} \real{z_1^* z_2}
\nonumber \\
f_2 = \frac{1}{2s} ( x_1 x_4 - x_2 x_3 ) = \frac{1}{2s} \im{z_1^* z_2}
\nonumber \\
f_3 = \frac{1}{4s} ( x_1^2 + x_2^2 - x_3^2 - x_4^2 ) =
\frac{1}{4s}\left(|z_1|^2 - | z_2|^2\right)
\; . \label{hopf}
\eea
Here we can identify $\br^4$ with $T^*\br^2$ with canonical coordinates
$x_1$, $x_3$ and momenta $p_1=\frac{x_2}{2s}$ and $p_2=\frac{x_4}{2s}$.

The SU(2)--action we are considering leaves the three--dimensional spheres
$S^3(R)=\{(x_1,x_2,x_3,x_4):x_1^2+x_2^2+x_3^2+x_4^2=R^2\}=\{(z_1,z_2):|z_1|^2
+|z_2|^2 =R^2\}$ invariant so that we can restrict the dynamics from the
full $\br^4$ to the submanifolds $S^3(R)$ on which the classical orbits
lie. In the following we choose $R=2s$ and work on $S^3(2s)$.

The functions (\ref{hopf}) and hence the Hamiltonian (\ref{hr}) have an
additional symmetry, being invariant under the action of U(1) given
by:
\begin{displaymath}
(z_1,z_2) \mapsto (e^{i\t} z_1 , e^{i\t}z_2 ) \;\;\;\; \t\in [0,2\p[
\; . \end{displaymath}
This allows us to project the Hamiltonian (\ref{hr}) from $S^3(2s)$ down
to the two-dimensional manifold $S^3(2s)/$U(1), which is homeomorphic to
the two-dimensional sphere. Indeed, the three functions
$(x_1,x_2,x_3,x_4) \mapsto S_j \equiv f_j(x_1,x_2,x_3,
x_4)$ ($j=1,2,3$) given in (\ref{hopf}) are the components of the
projection map from the three-sphere $S^3(2s)$ to the two-sphere $S^2(s)$
of the Hopf bundle U(1)$\rightarrow S^3\rightarrow S^2$ \ci{mor}.

On $S^2(s)$ the Hamiltonian (\ref{hr}) becomes simply $H= \sum_{j=1}^3
B_j S_j$ and therefore coincides with the Hamiltonian (\ref{ham}) for a spin in
a magnetic field. This shows that the latter is the reduction of a quadratic
Hamiltonian defined on a linear manifold {\footnote{To prove this rigorously,
we should show also that the restriction to $S^3(2s)$ of the symplectic
two-form $\bar{\O}$ is the pull-back of the symplectic two-form $\O$ we
defined on $S^2(s)$. The proof of this statement is straightforward, so
we will omit details here.}}.

\section{Coherent state path integrals for spin.}

Let us consider now the quantum mechanics of
a spinning particle, described by the Hamiltonian
\be
\hat H = \vec{B} \cdot \hat{\vec{S}} \label{qh} \; ,
\ee
where the spin operators $\hat{S}_j$ ($j=1,2,3)$ satisfy the usual
commuation relations $(\hbar=1)$:
\be
[\hat{S}_i,\hat{S}_j] = i \epsilon_{ijk} \hat{S}_k \; . \label{su}
\ee
We have already mentioned in the Introduction that the semiclassical
Bohr-Sommerfeld quantization turns out to be exact for this problem and
this seems to suggest that the SPA to the path integral for the
propagator and/or the partition function could be exact as well. In
addition, for such a simple problem, one can evaluate the partition
function exactly. Its expression
\be
\cz({\b}) = \mbox{Tr}\{ e^{-\b \hat H}\} = \frac{e^{\b Bs}}{1-e^{-\b B}} +
\frac{e^{-\b Bs}}{1-e^{\b B}} \label{qpf}
\ee
can be thought of as the sum of two terms, each corresponding to one of the
two poles of the sphere ($S_3=\pm s$), similarly to what happens
in the classical case.

In this Section we will briefly review the existing literature on the
evaluation of  the partition function for a quantum spin
in a magnetic field and in particular on the validity of the SPA applied
to this problem.

Let's begin by considering a set $\ket{l}$ of generalized coherent
states\ci{kla78,per}
labeled by one or more continuous variables that we denote collectively as $l$.
The $\ket{l}$'s will be assumed to be normalized. They are however
overcomplete, because although there is a resolution of the
identity associated with
them:
\be
1=\int dl \ket{l}\bra{l}
\lb{unity}
\ee
with ``$dl$" a suitable measure, in general they fail to be an orthonormal set:
$\scal{l}{l'}\neq 0$.

Here we are interested in the group SU(2), which is generated by the
spin-operator algebra (\ref{su}). Coherent states\ci{per} for a spin $s$
($2s$ being an integer) can be constructed as
\be
\ket{\theta,\phi}=e^{-i\phi\hat{S}_3}e^{-i\theta\hat{S}_2}\ket{0}
\; , \lb{def1}
\ee
where $\ket{0}$ denotes the highest-weight state of the spin-s
representation of SU(2) ($\hat{S}_3\ket{0}=s\ket{0}$) and $\t$, $\f$
are the two angular coordinates parametrizing $S^2$:
$0\leq\theta\leq\pi; 0\leq\phi\leq 2\pi$.
It is well known\ci{kla,per} that:
\be
\scal{\theta',\phi'}{\theta,\phi}=
\left[\cos\frac{\theta'}{2}\cos\frac{\theta}{2}e^{i(\phi'-\phi)/2}
+\sin\frac{\theta'}{2}\sin\frac{\theta}{2}e^{-i(\phi'-\phi)/2}\right]^{2s}
\label{25}\ee
and that the resolution of the identity associated with spin--coherent states
is:
\be
1=\frac{2s+1}{4\pi}\int_0^{2\pi}\int_0^{\pi}\ket{\theta,\phi}\bra{\theta,\phi}
\sin\theta d\theta d\phi \; .
\lb{compl}
\ee

Going back to the general case, let ${\hat H}$ be the Hamiltonian of a quantum
system. The matrix element of the propagator
\be
K(l_F,l_I,T)=\bra{l_F}e^{-i\hat{H}T}\ket{l_I}
\ee
can be represented as the following path integral:
\be
K(l_F,l_I,T)=\lim_{\eps\ra 0}\int\prod_{k=1}^N(dl_k)
\prod_{k=0}^N\bra{l_{k+1}}e^{-i\epsilon\hat{H}}\ket{l_k}
\; , \lb{pi}
\ee
where $\eps=T/N$ and
$\ket{l_0}\equiv\ket{l_I},\ket{l_{N+1}}\equiv\ket{l_F}$.
Eq.\rf{pi} can be rewritten as:
\be
K(l_F,l_I;T)=\lim_{\eps\ra 0}\int\prod_{k=1}^N(dl_k)\prod_{k=0}^N
e^{iA(l_{k+1},l_k)} \; ,
\lb{pi1}
\ee
where
\be
A(l_{k+1},l_k)=-i\ln\scal{l_{k+1}}{l_k}-\eps H(l_{k+1},l_k)
\ee
and
\be
H(l_{k+1},l_k)=
\frac{\bra{l_{k+1}}\hat{H}\ket{l_k}}{\scal{l_{k+1}}{l_k}} \; .
\ee
It is usually assumed that,
for $\eps$ small,
$\ket{l_{k+1}}$ is so close to $\ket{l_k}$ that one is allowed to
expand the former around the latter to leading order in $\eps$.  This leads to
\be
H(l_{k+1},l_k)\sim H(l_k)=\bra{l_k}{\hat H}\ket{l_k}
\lb{ca}
\ee
and to
\be
\ln\scal{l_{k+1}}{l_k}\sim\scal{\delta l_k}{l_k}
\; , \lb{caa}
\ee
where $\ket{\delta l_k}=\ket{l_{k+1}}-\ket{l_k}$. Hence the continuum
version of the path integral reads
\be
\int{\cal D}l \exp\left[ i\int_{0}^{T}dt \left[ i\scal{l}{\dot{l}}-H(l)
\right]\right]
\lb{b}
\ee
where $\ket{{\dot l}}= d\ket{l}/dt$.
This procedure is justified when the Hamiltonian contains an explicit kinetic
term that can act as a regulator concentrating the functional measure on
continuous paths.
This is the case of Wiener integral, where\ci{gel60,nel64}
the measure is concentrated on paths $q(t)$ satisfying the Lipschitz condition:
$|q(t+\eps)-q(t)|=O(\eps^{1/2})$.
The same holds true for the Feynman path integrals for massive particles in
not too singular potentials\ci{fey}.

In the case of spin, the Hamiltonian
(\ref{qh}) contains no regulators and the straightforward application of\rf{ca}
and\rf{caa} is
questionable. We will see indeed that it may lead to serious problems. Let's
be more specific. Taking again the magnetic field along the positive
z-axis we have, in terms of spin--coherent states\rf{def1}:
\be
\bra{l}{\hat H}\ket{l}\ra\bra{\theta,\phi}{\hat H}\ket{\theta,\phi}=
s B \cos\theta
\ee
and
\be
i\scal{l}{{\dot l}}\ra i\bra{\theta,\phi}\frac{d}{dt}\ket{\theta,\phi}
=s\cos\theta{\dot \phi} \; .
\ee
So we obtain the path integral:
\be
\bra{\theta_F,\phi_F}e^{-i\mu B\hat{S}_3T}\ket{\theta_I,\phi_I}=
\int {\cal D}\Omega
\exp\left[i\int_{0}^{T}\left[s\cos\theta\dot{\phi}-s\mu B\cos\theta
\right]dt\right] \; ,
\lb{d}
\ee
where $\cd\O = \sin\t\, \cd\t\, \cd \f$. From (\ref{d}) we can obtain the
canonical partition function $\cz(\b)$ by setting
$T=-i\beta$, $\ket{\theta_F,\phi_F}=\ket{\theta_I,\phi_I}$ and by tracing over
the angles. Following the standard procedure and Ref.\ci{nie88},
we have evaluated $\cz(\b)$ by time-slicing the
path integral and using the resolution of the identity\rf{compl} at each
intermediate (Euclidean) time.
Denoting with $\cz_N(\beta)$ the result for
$N$ time slices, we find for it without any further
approximation the expression\ci{biagio}:
\be
\cz_N(\beta)=\left(1+\frac{1}{2s}\right)^N \cz(\beta)\; ,
\lb{div}
\ee
where $\cz(\b)$ in the right hand side is the exact partition function
(\ref{qpf}){\footnote{In order to include paths turning around north pole
any number of times we need to extend, at every time slice,  the domain of
integration of the variable $\phi$ from $[0,2\pi]$ to
$[-\infty,+\infty]$. In fact without following such procedure one would
get, as in Ref.\ci{nie88}, a completely wrong partition function.}}.

As one can easily notice by simple inspection, in the limit
$N\ra\infty$, $\cz_N$
coincides with the exact partition function up to
a diverging prefactor. Such divergency disappears only in the classical
limit, namely
$\hbar\ra 0,s\ra\infty$ with $\hbar s={\rm const}$. On the other hand the
only approximations we used are those given by eqs.\rf{ca} and mainly\rf{caa}
which again are exact in the classical limit.
In the following of the paper we shall come back to this point.\\

We discuss now briefly two different approaches\ci{kla79,suz83}
that lead to the conclusion that in the case of a spinning particle in a
magnetic field the SPA to the path integral for the propagator
does indeed yield the exact result. This holds again up to a
normalization factor not taken into account in\cite{kla79,suz83}, since
the calculation is done in the continuum.

Klauder\ci{kla79} has proposed a modified form for the action that appears in
the path integral (\ref{b}), redefining the latter as:
\be
K(l_F,l_I;T)=
\lim_{\eps \ra 0}
\int{\cal D}l \exp\left[ i\int_{0}^{T}dt \left[ i\scal{l}{\dot{l}}
+\frac{1}{2}i\eps\bra{\dot{l}}(1-\ket{l}\bra{l})\ket{\dot{l}}-H(l)
\right]\right] \; .
\lb{mf}
\ee
The prescription here is that the limit for $\eps\ra 0$ should be taken
after having evaluated the path integral in the SPA.
For the case of a spin in a magnetic field,\rf{mf}  becomes:
\bea
& &K(\theta_F,\phi_F,\theta_I,\phi_I;T)= \label{i} \\
& &= \lim_{\eps\ra 0}\int \cd\O
\exp\left[i\int_{0}^{T}\left[s\cos\theta\dot{\phi}
+\frac{1}{4}is\eps(\dot{\theta}^2+\sin^2\theta\dot{\phi}^2)
-s\mu B\cos\theta\right]dt\right] \nonumber
\eea
and the action in\rf{i} becomes essentially that of a particle of charge
$s$ and mass $m=1/2 s \eps$ moving on the two-sphere $S^2$, coupled both
with a
magnetic monopole of unit strength located at the centre of the sphere and
with  a
constant electric-type field directed along the z-axis. In this context,
Dirac's quantization condition\ci{dir31} becomes identical with the spin
quantization condition, i.e.: $2s=$integer.

The classical equations of motion that can be derived from the action
in\rf{i}
are now second-order in time for $\eps\neq 0$, and hence are not plagued
by the already mentioned problem of overdetermination. Klauder has proved
that they can be solved explicitly and that the resulting
SPA to the path integral (or, better, what he calls the ``dominant SPA",
namely approximating the path integral
with $e^{iS_{cl}}$, without any prefactor originating from the
integration of gaussian fluctuations) is indeed exact. Actually, only the
free-spin case $B=0$ has been considered in Ref.\ci{kla79}, but the extension
to $B\neq 0$ is straightforward\ci{biagio}. It should be noted, incidentally,
that the term that Klauder has added to the action is in the form of a kinetic
energy term, thus providing the required regulator justifying the assumption
of continuously-varying paths.

The origin of the additional kinetic-type term, that looks somewhat
mysterious in Klauder's original paper\ci{kla78}, can be clarified as follows.
If we push the expansion of the logarithm in\rf{caa} one step further
beyond first order we obtain, with $\ket{\delta l}\sim\eps\ket{{\dot l}}
+\eps^2/2\ket{{\ddot l}}$:
\be
\ln\scal{l_{k+1}}{l_k}\sim
\eps\scal{\dot{l_k}}{l_k}+\frac{\eps^2}{2}\scal{\ddot{l_k}}{l_k}
-\frac{\eps^2}{2}(\scal{\dot{l_k}}{l_k})^2 \; ,
\ee
leading, in the continuum limit to
\be
\lim_{\eps\ra 0}\exp\left[ i\int_{t_i}^{t_f}dt\left[
i\scal{l}{\dot{l}} + \frac{1}{2}i\eps\left(-\scal{\ddot{l}}{l}
+\scal{\dot{l}}{l}\scal{l}{\dot{l}}\right)\right]\right]
\lb{h}
\ee
where we have used $\scal{l}{{\dot l}}=-\scal{{\dot l}}{l}$.
A final integration by parts of the term containing the second
derivative yields precisely Klauder's additional term.\\

Suzuki\ci{suz83} has adopted quite a different approach. Introducing two
additional resolutions of the identity, the propagator
\be
K(l_{F},l_I,T)=
\int_{l(0)=l_I}^{l(T)=l_F}{\cal D}l e^{i A_{FI}(l)}
\lb{r1} \; ,
\ee
where
\be
A_{FI}(l)=\int_{0}^T dt\left[ i\scal{l}{\dot{l}}-H(l)\right]
\lb{uu} \; ,
\ee
is rewritten in Ref.\ci{suz83} in the form:
\bea
K(l_{F},l_I,T)&=&\int\int dl_f dl_i
\scal{l_F}{l_f}\bra{l_f}e^{-i\hat{H}T}\ket{l_i}\scal{l_i}{l_I}\nonumber\\
&=&\int\int dl_f dl_i\scal{l_F}{l_f}\scal{l_i}{l_I}
\int_{l(0)=l_i}^{l(T)=l_f}{\cal D}l e^{i A_{fi}(l)}
\lb{r2} \; .
\eea
In general the parameter(s) $l$ labeling a coherent state can be defined in
terms of the expectation values of a suitable set of operators: position
and momentum for the free particle and the harmonic oscillator, spin
components in the case of spins. Taking then the latter as Cauchy data
for the canonical equations of motion one can determine how they evolve
classically in time, thus determining a ``classical" coherent state
$\ket{l(t)}$
as the coherent state labeled by $l(t)$, the evoluted at time $t$ of any given
initial parameter set $l$.
Then, Suzuki has argued that the SPA to the last
integral leads to
\be
\int_{l(0)=l_i}^{l(T)=l_f}{\cal D}l e^{i A_{fi}(l)}\sim
\delta(l_f-l_i(T))e^{iA_{fi}}
\lb{sc} \; ,
\ee
where $l_i(T)$ is the evoluted at time $T$ of $l_i$, and that the final result
of the application of the SPA is the semiclassical propagator
\be
K_{sc}(l_F,l_I,T)=\int dl_i\scal{l_F}{l_i(T)}
\scal{l_i}{l_I}e^{iA_{fi}}
\lb{mir} \; ,
\ee
where $A_{fi}$ is the classical action evaluated along the classical path
leading from $l(0)=l_i$ to $l(T)$. Once again, the semiclassical
propagator\rf{mir} yields the exact result for the propagator of a
spinning particle.

We would like to show  here that the possible exactness of Eq.\rf{mir} has very
little to do with the path integral formalism itself, and that it follows
rather from a single assumption, one that amounts basically to assuming
that the Ehrenfest theorem be applicable in the present case.
Let ${\hat l}$ be the set of operators whose expectation values label a
given coherent state.
To each coherent state $\ket{l}$ we can associate two different time
dependent states:\\
-- the ``classically'' time evoluted state $\ket{l(t)}$ defined above and
therefore such that
\be
 \bra{l(t)}{\hat l}\ket{l(t)}= l_{cl}(t)    \; ,
\ee
$l_{cl}(t)$ being the classical trajectory,\\
-- the quantum time evoluted state $\ket{l,t}$ obtained through the
application of the full quantum evolution operator:
\be
\ket{l,t}=e^{-i{\hat H} t}\ket{l}\; .
\ee
We require now the expectation values on the quantum evoluted state to
evolve in time according to the classical equations of motion:
\be
\bra{l,t}{\hat l}\ket{l,t}=l_{cl}(t)\; .
\ee
In other words we are assuming, as anticipated,
the validity of the Ehrenfest theorem. A sufficient condition for this to
happen is that the Heisenberg equations of motion be linear equations.
This is certainly true for quadratic Hamiltonians with conventional
canonical coordinates and momenta and the standard Poisson brackets
among them. For spins, the Hamiltonian (\ref{sh}) is not of the quadratic type,
but the Poisson brackets (\ref{pb}) lead nonetheless to linear equations of
the motion and hence to the validity of the Ehrenfest theorem.
If we now assume that the set of operators ${\hat l}$
act irreducibly on the Hilbert space of states (which is true,
e.g., for harmonic oscillators and for spin systems)
the two states $\ket{l,t}$ and $\ket{l(t)}$ will differ
at most by a phase, i.e.:
\be
\ket{l,t}=e^{i\chi(t)}\ket{l(t)}
\; . \lb{forte}
\ee
Differentiating with respect to time we obtain
\be
\bra{l(t)}\frac{d}{dt}\ket{l(t)}=-i\dot{\chi} + \bra{l,t}\frac{d}{dt}
\ket{l,t}
\lb{emb}
\ee
and hence
\be
 \scal{l}{\dot{l}}_{cl}=-i\dot{\chi} - i\bra{l,t}\hat{H}\ket{l,t} \; .
\ee
If, as it is presently the case, the Hamiltonian does not contain time
derivatives, we obtain furthermore:
\be
\dot{\chi}=i \scal{l}{\dot{l}}_{cl} - \bra{l(t)}\hat{H}\ket{l(t)}=
i\scal{l}{\dot{l}}_{cl} - H(l(t))
\ee
and eventually, integrating the last equation and up to an irrelevant constant
phase:
\be
\chi(t)=\int_0^t\left[ i\scal{l}{\dot{l}}_{cl}-H(l)\right] dt'
\lb{mia}\; .
\ee
The r.h.s. of\rf{mia} is exactly the classical action, hence
$\chi(t)=A_{cl}(t)$.
Then Suzuki's result follows at once from:

\newpage

\bea
K(l_{F},l_I,T)&=&\bra{l_F}e^{-i\hat{H}T}\ket{l_I} \nonumber\\
&=&\int dl_i \bra{l_F}e^{-i\hat{H}T}\ket{l_i}\scal{l_i}{l_I}\nonumber\\
&=& \int dl_i \scal{l_F}{l_i(T)}\scal{l_i}{l_I}e^{i\phi(T)}\nonumber\\
&=& \int dl_i \scal{l_F}{l_i(T)}\scal{l_i}{l_I}e^{iA_{fi}}
\lb{rid}\; ,
\eea
which is the desired result.

Let's analyze\rf{rid} in the specific case of a spinning particle.
As already pointed out the Hamiltonian\rf{sh} leads to linear Heisenberg
equations of motion and hence to the validity of formula\rf{rid}.
It seems to us that this is the ultimate reason for the apparently surprising
fact that SPA to path integrals and/or
semiclassical quantization leads to the exact result for the quantum
propagator of a spinning particle.
We would like to stress here that this is also true for the motion in a
magnetic field with an arbitrary dependence on time. Formula\rf{rid} reduces
therefore the quantum problem of the calculation of the spin propagator in an
arbitrary magnetic field to the solution of the classical equation of motion.

The expression for the propagator can of course be worked out explicitly only
when the classical equation of motion can be solved analytically.
With a constant magnetic field along the z-axis, formula\rf{mir} reads:
\be
K(\theta_F,\phi_F,\theta_I,\phi_I;T)=
\frac{2s+1}{4\pi}\int d\Omega_i
\scal{\theta_F,\phi_F}{\theta_i(T),\phi_i(T)}
\scal{\theta_i,\phi_i}{\theta_I,\phi_I}e^{iA_{cl}} \; .
\ee
Since the solutions of the classical equations of motion are
$\theta_i(T)=\theta_i ;
\phi_i(T)=\phi_i + \mu B T$, while $A_{cl}=0$, the above expression
becomes:
\be
K(\theta_F,\phi_F,\theta_I,\phi_I;T)=
\frac{2s+1}{4\pi}\int d\Omega_i
\scal{\theta_F,\phi_F}{\theta_i,\phi_i+\mu B T}
\scal{\theta_i,\phi_i}{\theta_I,\phi_I}
\lb{msp}\; .
\ee
The integral in\rf{msp} can be done easily by noting that
\linebreak
$\scal{\theta_F,\phi_F}{\theta_i,\phi_i + \mu B T}=
\scal{\theta_F,\phi_F - \mu B T}{\theta_i,\phi_i}$ and using the
resolution of the identity. We obtain eventually
\be
K(\theta_F,\phi_F,\theta_I,\phi_I;T)=
\scal{\theta_F,\phi_F-\mu B T}{\theta_I,\phi_I} \; ,
\ee
which indeed gives the exact propagator, upon using formula\rf{25} for the
overlap of two coherent states.

\section{The complex saddle point.}

We devote this Section to another approach to path integral for
spin. It has been used for example in\ci{vak91} and in\ci{rks},
where however only continuous and hence formal expressions for the
action and the path integral have been studied.

Here we will study the path integral {\it discrete} version for
the propagator (\ref{b}). The  method is based on a
different choice of the spin--coherent states. Instead of (\ref{def1}),
we consider the following states, labeled by a complex parameter $\m$:
\be
\ket{\mu}=\frac{e^{\mu S^-}}{(1+|\mu|^2)^s}\ket{0} \; , \label{cohe}
\ee
where we have set $\hat{S}^{\pm} = \hat{S}_1 \pm i\hat{S}_2$. To be more
precise, $\m$ parametrizes the homogeneous space SU(2)/U(1) $= S^2$
by means of the stereographic projection of $S^2$ onto $\br^2$.
With respect to the spherical coordinates $\theta$ and $\phi$
used previously, one has
$\m = \mbox{tan} \frac{\t}{2} e^{i\f}$ and it is also not difficult to check
that the state (\ref{cohe}) coincides up to a phase with the state
$\ket{\t,\f}$ defined in\rf{def1}.

To write down an explicit expression for the path integral in terms of
the coherent states (\ref{cohe}), we need the matrix
elements
\bea
&& \scal{\lambda}{\mu}= \frac{(1+\lambda^*\mu)^{2s}}{(1+|\mu|^2)^s
(1+|\lambda|^2)^s}
\lb{scal} \\
&& \bra{\lambda}{\hat S}_3\ket{\mu}=\scal{\lambda}{\mu} s \frac{1-
\lambda^*\mu}{1+\lambda^*\mu} \nonumber \\
&& \scal{\mu}{{\dot \mu}}=\frac{s}{1+|\mu|^2}
({\dot \mu}\mu^*-{\dot\mu}^*\mu) \nonumber
\eea
and the explicit formula for the resolution of the identity:
\be
\frac{2s+1}{\p} \int \frac{d^2\m}{(1+|\m|^2)^2} \ket{\m} \bra{\m} =1
\label{comple}
\ee
where $d^2\m = d(\real\m) d(\im\m)$. After some algebra, one finds that
the path integral representation of the propagator is
given by
\be
K(\m_F^*,\m_I;T) = \int_{\begin{array}{c} \m (0)=\m_I \\
\bigl(\m (T)\bigr)^*=
\m_F^*\end{array}} \frac{d^2\m}{(1+|\m|^2)^2} e^A \label{prop}\; ,
\ee
where the action for a spin in a magnetic field is
\be
A= is \int_0^T dt
\left( i \frac{\dot{\m} \m^* - \m \dot{\m}^*}{1+|\m|^2} - B \frac{1-
|\m|^2}{1+|\m|^2} \right)  \;. \label{act}
\ee

Again the SPA equations that are derived by extremizing
(\ref{act}):
\be
\left\{\begin{array}{l}
{\dot \mu}= i B \mu\\
{\dot \mu^*}= -i B \mu^*
\end{array} \right.
\lb{aa}
\ee
are first-order, so that they do not admit of a solution for boundary
conditions of the form:
\be
\left\{\begin{array}{l}
 \mu (0)= \mu_I\\
 \bigl(\mu (T)\bigr)^*= \mu_F^*
\end{array} \right.
\lb{bb}
\ee
with arbitrary $\mu_I$ and $\mu_F$.

This problem does however admit of solutions\ci{vak91,fad76}
if we enlarge our variable
space from $\bc$ to $\bc^2$ and look for saddle point solutions
for which $\m(t)$ and $\m^*(t)$ are {\it not} complex-conjugate one of
the other. In other words, we have to consider $\m$ and $\m^*$ as
independent complex variables and look for solutions of (\ref{aa})
satisfying the boundary conditions
\be
\left\{\begin{array}{l}
\m(0) = \m_I\\
\m^*(T) = \m_F^* \end{array} \right. \lb{bb1}\; .
\ee
These so called complex saddle point solutions are easily found to be
\be
\left\{\begin{array}{l}
 \mu(t)= e^{i B t}\mu_I\\
 \mu^*(t)= e^{i B(T-t)} \mu^*_F
\end{array} \right.
\lb{cc}
\ee
and, to complete the SPA, one has further to evaluate the contributions
of the gaussian fluctuations around the solutions\rf{cc}.

This has been done by the authors of\ci{vak91,rks}, who have
concluded
that one can obtain the exact propagator in this way. They have however
worked with the continuous expression (\ref{act}) and therefore they
have performed only formal calculations.
Funahashi {\em et al}.\ci{fun94} have considered the discrete version of the
integral but they have applied SPA to the calculation of the partition
function only. This problem, as pointed out by the same authors, is
simpler than the calculation of the propagator since the
overdetermination problem can be overcome without enlarging the variable
space from $\bc$ to $\bc^2$.

Our aim in this Section is to examine whether the SPA to the path
integral (\ref{prop}) is exact by performing the calculation in the
{\it discrete} version of the path integral. The latter is given by\rf{pi1},
where now the generic index $l$ stands for the complex parameter $\m$.
By using\rf{scal} and rearranging the terms, one eventually gets:
\be
K(\mu_F^*,\mu_I,T)=
{\cal N}\int
\prod_{j=1}^{N-1}\frac{d^2\mu_j}{(1+\mu_j^*\mu_j)^2}
e^{A}
\lb{iop}
\ee
with
\be
{\cal N} = \left(\frac{2s+1}{\p}\right)^{N-1}
\frac{1}{(1+|\m_F|^2)^s (1+|\m_I|^2)^s} \label{nf} \; ,
\ee
whereas the discretized action $A$ is given by
\be
A=\sum_{j=1}^{N}\left[ A_{j,j-1}+ H_{j,j-1} \right]+
2s \ln(1+\mu_F^*\mu_{N}) \label{da}
\ee
\be
A_{j,j-1} \equiv 2s\ln \frac{1+\mu_j^*\mu_{j-1}}{1+\mu_j^*\mu_j}
\lb{azf}
\ee
\be
H_{j,j-1}=-iBs\eps\frac{1-\mu_j^*\mu_{j-1}}{1+\mu_j^*\mu_{j-1}}\; .
\ee

The expression for $A_{j,j-1}$ given in \rf{azf} comes from the
exponentiation of the overlaps $\scal{\m_j}{\m_{j-1}}$ ($j=1,\cdots,N$).
Let us remark that we have {\it not} expanded
the overlaps to first order in the difference $\ket{\m_j} - \ket{\m_{j-
1}}$, as usually done to recover the continuum version (see (\ref{caa})).
As already pointed such an expansion would
be correct only in the presence of a regulating term in the action.
Thus we will work only with (\ref{azf}) and we will proceed
now to evaluate the SPA to the multidimensional integral (\ref{iop}),
according to the formula
\be
K(\mu_F^*,\mu_I,T)_{sc}=
{\cal N}\prod_{j=1}^{N-1}\left(\frac{1}{(1+\mu_j^*\mu_j)^2}\right)_{ cl}
e^{A_{cl}}\frac{\pi^{N-1}}{\left(\det M_N\right)^{1/2}}
\lb{scl}\; ,
\ee
where $M_N$ is the $2(N-1)\times 2(N-1)$ matrix of quadratic
fluctuations around the classical solutions of the discretized action
(\ref{da}):
\be
A= A_{cl} +
(\delta\mu_1^*,\delta\mu_1,\ldots,\delta\mu_{N-1}^*,\delta\mu_{N-1}) M_N
\mbox{$\left(\begin{array}{c}
\delta\mu_1\\
\delta\mu_1^*\\
\vdots\\
\delta\mu_{N-1}\\
\delta\mu^*_{N-1}
\end{array}\right)$} \; .
\ee
The subscript ``$cl$" means that the function (the action in this case)
has to be evaluated on the classical solution, which satisfies the
saddle point equations:
\be
\left\{\begin{array}{ccc}
\frac{\mu_{j+1}^*-\mu_j^*}{1+\mu_j^*\mu_j}&=&-i\eps B\frac{\mu_{j+1}^*}
{1+\mu_{j+1}^*\mu_j}\\
\frac{\mu_{j}-\mu_{j-1}}{1+\mu_j^*\mu_j}&=&+i\eps B\frac{\mu_{j-1}}
{1+\mu_{j}^*\mu_{j-1}}
\end{array}\right.
\lb{gu}\; .
\ee
In (\ref{gu}), as well as in (\ref{da}), we have explicitely written
$|\m_j|^2$ as $\m_j \m_j^*$ to stress the fact that in the search of
the saddle point solutions we have to treat $\m_j$ and $\m_j^*$ as
independent complex variables. Indeed, exactly as in the continuum, the
equations (\ref{gu}) are incompatible with the boundary conditions
\be
\left\{\begin{array}{l}
\mu_0=\mu_I\\
\mu^*_N=\mu^*_F
\end{array}\right.
\lb{bou}
\ee
unless $\m_j$ and $\m_j^*$ are treated as independent variables.

We know that the classical solution has to fulfill the property $\m_{j+
1}|_{cl} = \m_j|_{cl} + O(\e)$, so that we can approximate the
denominators in the right hand side of ({\ref{gu}), to get
\be
\left\{\begin{array}{ccc}
\frac{\mu_{j+1}^*-\mu_j^*}{1+\mu_j^*\mu_j}&=&-i\eps B\frac{\mu_{j}^*}
{1+\mu_{j}^*\mu_j} + O(\eps^2)\\
\frac{\mu_{j}-\mu_{j-1}}{1+\mu_j^*\mu_j}&=&+i\eps B\frac{\mu_{j-1}}
{1+\mu_{j}^*\mu_{j}} + O(\eps^2)
\end{array}\right. \; .
\ee
To order $\e$, we find for the solutions satisfying the
boundary conditions (\ref{bou}) and for the classical action the
expressions:
\be
\left\{\begin{array}{l}
\mu_{j}=(1+i\eps B)^j\mu_I\\
\mu_{j}^*=(1+i\eps B)^{N-j}\mu_F^*
\end{array}\right.
\ee
\be
 A_{cl} = -i s B T \label{clact} \; .
\ee

To complete the calculation of the right hand side of (\ref{scl}) we
have to compute the determinant of the gaussian fluctuation matrix $M_N$.
 As for the solutions of the classical equations (\ref{gu}), we
evaluate the matrix elements to the order $O(\e)$. Neglecting therefore
all the terms of at least order $O(\e^2)$ we find that the only non-zero
matrix elements are
\bea
a_j&&\equiv\left.\frac{\pa^2 A}{\pa\mu_j^*\pa\mu_j} \right|_{cl}=
\left.\frac{\pa^2 A}{\pa\mu_j\pa\mu_j^*} \right|_{cl}\nonumber\\
& &=\left.\frac{-2 s}{(1 +\mu_j^*\mu_j)^2}\right|_{cl}=
\frac{-2 s}{(1 +(1+i\eps B)^N\mu_F^*\mu_I)^2} \; ,\\
b_j&&\equiv\left.\frac{\pa^2 A}{\pa\mu_j\pa\mu_{j+1}^*} \right|_{cl}=
\left.\frac{\pa^2 A}{\pa\mu_{j+1}^*\pa\mu_{j}} \right|_{cl}\nonumber\\
& &=\frac{2 s}{1+\mu_{j+1}^*\mu_j}\left[
\frac{1}{1+\mu_j^*\mu_j}+i\eps B\frac{1}{(1+\mu_{j+1}^*\mu_j)^2}
\right]_{cl}\nonumber\\
& &=\frac{2s (1+ i\eps B)}{(1+(1+i\eps B)^N \mu_I\mu_F^*)^2}\; ,
\eea
Defining
\be
A_i=\left(\begin{array}{cc}
a_i&0\\
0&a_i\end{array}\right) \;\;
B_i=\left(\begin{array}{cc}
0&0\\
0&b_i\end{array}\right) \;\;
C_i=\left(\begin{array}{cc}
b_i&0\\
0&0\end{array}\right) \; ,
\ee
the matrix $M_N$ is given by
\be
M_N=\left(
\begin{array}{cccccc}
A_1&B_1& & & & \\
C_1&A_2&B_2& & & \\
 & & &\ddots& & \\
 & & &C_{N-3}&A_{N-2}&B_{N-2}\\
 & & & &C_{N-2}&A_{N-1}\end{array}\right) \; .
\ee
Thus we have
\be
\det M_N = a_1^2\ldots a_{N-1}^2= \left[
\frac{-2 s}{(1+ (1+i\eps B)^N\mu_I\mu_F^*)^2}\right]^{2(N-1)}\; ,
\ee
which can be finally inserted, together with (\ref{clact}), in
(\ref{scl}), yielding:
\be
K(\mu_F^*,\mu_I,T)_{sc}=\left(1+\frac{1}{2s}\right)^{N-1}
\frac{e^{-i s B T}(1+\mu_I\mu_F^*(1+ i B T/N)^N)^{2s}}
{(1+|\mu_F|^2)^s(1+|\mu_I|^2)^s} \label{end}\; .
\ee
This is again the expected result up to the divergent normalization
factor $(1+1/2s)^{N-1}$. Indeed since $\lim_{N\ra\infty} (1+ i B T/N)^N=
e^{iBT}$, but for the prefactor, we obtain
\be
K(\mu_F^*,\mu_I,T)_{sc}=
\frac{e^{-i s B T}(1+\mu_I\mu_F^* e^{ i B T})^{2s}}
{(1+|\mu_F|^2)^s(1+|\mu_I|^2)^s}\; ,
\lb{pre}
\ee
which coincides with the exact propagator.

\section{Conclusions.}
In this paper we have examined the problem of a spin in a magnetic field.
We have seen that the SPA applied to the calculation of the classical
partition function yields the correct result. In quantum mechanics the
situation is more complicated. We have reexamined the different
approaches that have been used in literature to prove the exactness of
SPA in the calculation of the quantum propagator and partition function.
All these methods, in particular those proposed by Klauder\ci{kla79} and
by Keski-Vakkuri {\em et al}.\ci{vak91}, make use of the continuum expression
of the path integral and hence reproduce the correct result only
formally.

To test the validity of the SPA for a spin system we decided to work with
the very definition of the path integral, namely with its discrete version.
We have written the discrete path integral for the partition function
and the propagator in terms of spin coherent states and then considered
two different kinds of approximations. In Section III we have expanded the
overlaps $\scal{l_k}{l_{k-1}}$ to first order in $\eps=T/N$ and then
performed an exact integration. On the contrary, in Section IV we have
made no expansion for the overlaps and applied instead the complex
saddle point method to evaluate the SPA of the path integral.
In both cases, we have found that the exact result is reproduced
correctly only up to an infinite normalization factor,
$\lim_{N\ra\infty}(1+1/2s)^N$, which goes to 1 in the classical limit
$s\ra\infty$, provided the latter is performed first.
Notice that this fact is not a drawback just of the
spin Hamiltonian. Indeed one can repeat easily the calculation for
$H\equiv 0$ and get the same infinite constant.

All this seems to suggest that, for those Hamiltonians that do not
contain a regulating term (such as a kinetic part), any
approximation that restricts the class of quantum paths in phase space,
by imposing some regularity conditions on them, yields a wrong result.
In our case this shows up in the appearance of an infinite prefactor.

We would like to conclude by commenting on the paper by Funahashi et
al.\ci{fun94}, who have been able to reproduce the exact partition function
(with the correct prefactor) by performing an SPA in the discrete path
integral, written again in terms of spin coherent states.
They have obtained this result by taking into account also the gaussian
fluctuations coming from the factor $1/(1+ |\mu|^2)^2=\sin\theta/4$
appearing in the integration measure.
We do not want to explain this technique which is described in
detail in\ci{fun94}. Here we notice only that the inclusion of fluctuations
coming from the measure induces a shift in the multiplicative factor
appearing in front of the action from $2s$ to $2s+1$.
Effectively, we can say that such a method amounts to choosing
$2s+1$ as parameter for a semiclassical expansion.
If so, we should not evaluate the path integral following\rf{scl}, but
according to:
\be
K(\mu_F^*,\mu_I,T)_{sc}=
{\cal N}\prod_{j=1}^{N-1}\left(
\frac{ e^{-\frac{{\widetilde A}}{N-1} } }{(1+\mu_j^*\mu_j)^2}\right)_{ cl}
\frac{\pi^{N-1}e^{(2s+1){\widetilde A_{cl}}}}
{\left(\det {\widetilde M_N}\right)^{1/2}}
\lb{last}\; ,
\ee
where ${\widetilde A}=A/2s$ and ${\widetilde M_N}$ is the matrix of
gaussian fluctuations of $(2s+1){\widetilde A}$ around the classical solution,
so that
\be
\det {\widetilde M_N}= a_1^2\ldots a_{N-1}^2= \left[
\frac{-2(s+1)}{(1+ (1+i\eps B)^N\mu_I\mu_F^*)^2}\right]^{2(N-1)}
\; .\label{ult}
\ee
It is exactly the factor $2s+1$ in\rf{ult} which cancels the same
factor in ${\cal N}$, yielding the correct propagator.

This is quite a remarkable result. In our opinion, however, the derivation
presented in\ci{fun94} needs some clarification. To us it seems to be
inconsistent to include gaussian fluctuations of the measure factor in
the calculation of the SPA, without considering also its contributions to the
saddle point equations.
In\ci{fun94} this method works only because the first derivatives
of the measure factor {\em do} vanish at the classical solution, which in
this case correspond to $\mu=0$ or $\mu=\infty$ ($\theta=0,\pi$ in
angular coordinates). But this would not be the case in slightly more
complicated situations, for example for the calculation of the
propagator or when considering complex saddle point solutions.

\appendix
\section{The Duistermaat-Heckman theorem.}

\setcounter{equation}{0}
\renewcommand{\theequation}{A\arabic{equation}}

Let us consider an oscillatory integral of the kind
\be
I(t)\equiv \left( \frac{t}{2\p} \right)^n \int_M \s e^{itf} \label{osc}
\ee
over a (2n)-dimensional manifold $M$ with volume form $\s$. If $M$ is a
Riemannian manifold, under rather mild hypotheses, namely that the function $f$
be a Morse function, i.e. that the Hessian matrix of $f$ be non-singular at
all critical points of $f$ ($\det\mbox{Hess}_P(f)\neq 0$
if $\nabla f(P)=0$), it is possible to show\ci{gs} that for large values
of the parameter $t$, one has
\be
I(t) = \sum_P c_P e^{itf(P)} + O(t^{-1}) \label{spa}\; ,
\ee
where the sum ranges over all critical points of $f$ and the
coefficients are given in terms of the determinant of the gaussian
fluctuations of $f$ around the critical points:
\be
c_P = \exp\left[ i \frac{\p}{4} \mbox{sgn}\mbox{Hess}_P(f)\right] \left[
\det\mbox{Hess}_P(f)\right]^{-\frac{1}{2}} \label{coeff} \; .
\ee
Here the signature $\mbox{sgn}A$ of a symmetric real-valued
nonsingular matrix $A$ is
defined as the number of its positive eigenvalues minus the number of
its negative eigenvalues.

Everybody is familiar with the elementary result that the remainder term
$O(t^{-1})$
vanishes identically if $M$ is the linear manifold $\br^{2n}$ with volume
form $\s=dx_1 \cdots dx_{2n}$ and the function $f$ is a quadratic form:
$f = \frac{1}{2} Q \vec{x}\cdot \vec{x}- \vec{\xi}\cdot \vec{x}$, $Q$
being any symmetric real-valued (2n)-dimensional non-singular matrix. In
this case the only critical point of $f$ is $\vec{x}_0=Q^{-1}\vec{\xi}$
and $\mbox{Hess}_{\vec{x}_0}(f)=Q$, so that (\ref{spa}) with $O(t^{-1})\equiv
0$  gives simply the formula for a gaussian integral.

The theorem of Duistermaat-Heckman\ci{dh} and
its generalization due to Berline
and Vergne\ci{bv}
establish under which conditions an integral of the kind
(\ref{osc}) can be exactly evaluated in the SPA, i.e. when $O(t^{-1})\equiv 0$.

Let $M$ be a compact (2n)-dimensional manifold with symplectic two-form
$\O$ and suppose $M$ is acted upon by a compact Lie group $G$, whose
action is symplectic and Hamiltonian\ci{gs}. Let us denote with $\chi_{\eta}$
the fundamental
vector field on $M$ generated by the action of the element $\eta$ in
the Lie algebra $\bar{G}$ of $G$ and with $f_{\eta}$ the associated
Hamiltonian function ($i_{\chi_{\eta}}\O = df_{\eta}$, where $i$ denotes
the contraction). Then, if $\chi_{\eta}$ is non-degenerate, i.e. if it
is zero only at the fixed points of $G$, the following results hold\ci{gs}:\\
1) $f_{\eta}$ is a Morse function;\\
2) \be
\left( \frac{t}{2\p} \right)^n \int_M \frac{\O^n}{n!} e^{itf_{\eta}}
= \sum_P c_P e^{itf_{\eta}(P)}  \label{dh}
\ee
where the the sum ranges over the critical points of $f$, i.e. over
the points $P$ such that $\chi_{\eta}(P) =0$ and the coefficients $c_P$
are given by
\be
c_P = \frac{i^n}{\l_1 \l_2 \cdots \l_n} \; , \label{co1}
\ee
the $\l_j$'s being the coefficients appearing in the matrix $L_P$ of the
derivatives at $P$ of the components of the vector field $\chi_{\eta}$,
$[L_P]^{ij}=\left(\frac{\partial\chi_{\eta}^i}{\partial x_j}\right)_P$,
which in a
suitable {\it positively oriented} basis can always be written as
{\footnote {If $M$ is equipped with a Riemannian metric, formula
(\ref{co1}) is seen to coincide with (\ref{coeff}).}}
\begin{displaymath}
L_P = \left[ \begin{array}{ccccccc} 0&\l_1& & & & & \\ \l_1&0& & & & & \\
 & &0&\l_2& & & \\ & &-\l_2&0&  & & \\  & & & & \ldots & & \\
 & & & & & 0&\l_n \\ & & & & & -\l_n & 0 \end{array} \right] \; .
\end{displaymath}

{}From a physical point of view, (\ref{dh}) can be applied to classical
statistical mechanics for the computation of the partition function:
\be
{\cal Z} = \int_M \frac{\O^n}{n!} e^{-\b H} \label{pf2}\; .
\ee
In this case the vector field $\chi_{\eta}$ is given by the Hamiltonian
vector field $\D_H$, where now $H$ plays the role of the function
$f_{\eta}$. The assumptions of the Duistermaat-Heckman theorem require
$\D_H$ to be non-degenerate and to be the
fundamental vector field associated with an element $\eta$ of the Lie
algebra $\bar{G}$ of a compact Lie group $G$ acting symplectically on
the manifold $M$. If these two conditions holds, we can apply formula
(\ref{dh}) to conclude that (we have set $\b=-it$):
\be
{\cal Z}(\b) = \left( \frac{2\p}{i\b} \right)^n \sum_P c_P e^{-\b H(P)}
\label{exact}\; ,
\ee
where\\
i) the sum ranges over the stationary points of the Hamiltonian,
$\D_H(P)=0$;\\
ii) the coefficients $c_P$ are given by (\ref{co1}) in terms of the
$\l_j$'s, the latter
being the coefficients of the matrix $[L_P]_{ij} =
\left(\frac{\partial \D_H^i}{\partial x_j}\right)_P$,
written in a suitable positive
oriented basis.\\

Let us go back now to the problem of a spin in a magnetic field. We want
to show that the Duistermaat-Heckman theorem can be applied to this
problem, so that the {\it exact} partition function can be calculated by
means of formula (\ref{exact}). The Hamiltonian vector field associated
to the Hamiltonian (\ref{ham}) is
\be
\D = B \left( -S_2 \frac{\partial}{\partial S_1} + S_1
\frac{\partial}{\partial S_2} \right) \label{hvf}
\ee
and it is easy to recognize that it is proportional to the generator
$\chi_{\eta}= \frac{\partial}{\partial \f}$ of the rotations about the third
axis (i.e. the axis along the constant magnetic field). Thus the Lie
group that acts symplectically on the phase space manifold $S^2(s)$ is
simply given by $U(1)$ in this case.

To apply (\ref{exact}) we have first of all to find the critical
points of the Hamiltonian, which are given by the North and the South
poles of the sphere:
\be
P_{\pm} \equiv (0,0,\pm s) \label{sta}
\ee
and then to compute the coeffcients $c_{P_{\pm}}$ according to (\ref{co1}).
In the tangent space of $P_+$ and $P_-$ we choose to work with the
positevely oriented basis $\left( \frac{\partial}{\partial S_1},
\frac{\partial}{\partial S_2}\right)$ and $\left(
\frac{\partial}{\partial S_2}, \frac{\partial}{\partial S_1}\right)$
respectively. With
respect to these bases
\be
L_{P_{\pm}}(\D) = \mp B \left[ \begin{array}{cc} 0&1\\-1&0 \end{array}
\right] \label{mat} \; ,
\ee
so that
\be
c_{P_{\pm}} = \mp \frac{i}{B} \label{c}\; .
\ee
We can finally compute the partition function in the SPA as
\be
{\cal Z} = \frac{2\p}{\b B} \left( e^{\b B s} - e^{-\b B s} \right)
\label{sfpf}\; ,
\ee
which, in agreement with the Duistermaat-Heckman theorem, coincides with
the exact partition function (\ref{pfs}) for a spin in a magnetic field.


\begin{thebibliography}{99}
\bibitem{fey}
R. P. Feynman, A. R. Hibbs, {\em Quantum Mechanics and Path
Integrals} (McGraw-Hill, New York, 1965).
\bibitem{schu68}
L. S. Schulman, Phys. Rev. {\bf 176},1558 (1968).
\bibitem{schu}
L. S. Schulman, {\em Techniques and Applications of Path Integrals}
(John  Wiley and Sons, New York, 1981).
\bibitem{kla78}
J. R. Klauder, in {\em Path Integrals}, Proceedings of the NATO Advanced
Summer Institute, G. J. Papadopoulos and J. T. Devreese eds. (Plenum, New
York, 1978).
\bibitem{kla79}
J. R. Klauder, Phys. Rev. D{\bf 19},2349 (1979).
\bibitem{kla}
J. R. Klauder, B. S. Skagerstam, {\em Coherent States, Applications in
Physics and Mathematical Physics} (World Scientific, Singapore, 1985).
\bibitem{per}
A. Perelomov, {\em Generalized Coherent States and their Applications}
(Springer-Verlag, Berlin, 1986).
\bibitem{nie88}
H. B. Nielsen, D. R\"ohrlich, Nucl. Phys. B {\bf 299},471 (1988).
\bibitem{sto89}
M. Stone, Nucl. Phys. B {\bf 314},557 (1989).
\bibitem{joh89}
K. Johnson, Ann. Phys. {\bf 192},104 (1989).
\bibitem{bal}
A. P. Balachandran, G. Marmo, B. S. Skagerstam and A. Stern, {\em Gauge
symmetries and fibre bundles} (Springer-Verlag, Berlin, 1983).\\
A. P. Balachandran, G. Marmo, B. S. Skagerstam and A. Stern, {\em Classical
Topology and Quantum States} (World Scientific, Singapore, 1991).
\bibitem{biagio} P. Pieri, Ph.D. Thesis, Bologna University, unpublished.
\bibitem{dir}
P. A. M. Dirac, {\em Lectures on Quantum Mechanics} (Belfer Graduate School
of Science, New York, 1964).
\bibitem{vak91}
E. R. Keski-Vakkuri, A. Niemi, G. Semenoff, O. Tirkkonen, Phys.
Rev. D {\bf 44},3899 (1991).
\bibitem{suz83}
T. Suzuki, Nucl. Phys. A {\bf 398},557 (1983).\\
H. Kuratsuji, T. Suzuki, Prog. Theor. Phys. Suppl. {\bf 74} and {\bf 75},209
(1983).\\
T. Fukui, J. Math. Phys. {\bf 34},4455 (1993).
\bibitem{dh}
J. H. Duistermaat and G. J. Heckman, Invent. Math. {\bf 69},259 (1982) and
{\bf 72}, 153 (1983).
\bibitem{bv}
N. Berline and M. Vergne, Duke Math. Jour. {\bf 50},539 (1983).
\bibitem{mssv}
G. Marmo, S. G. Saletan, A. Simoni and B. Vitale, {\em Dynamical Systems}
(John Wiley and Sons, New York, 1985).
\bibitem{glmv}
J. Grabowski, G. Landi, G. Marmo and G. Vilasi, Fortschr. der Phys.
{\bf 42},5 (1994).
\bibitem{amr}
R. Abraham, J. E. Marsden, T. Ratiu, {\em Manifolds, Tensor analysis and
Applications} (Springer-Verlag, Berlin, 1988).
\bibitem{mor}
G. Morandi, {\em The Role of Topology in Classical and Quantum Physics},
(Springer-Verlag, Berlin, 1992).
\bibitem{gel60}
M. Gelfand, A. M. Yaglom, J. Math. Phys. {\bf 1},48 (1960).
\bibitem{nel64}
E. Nelson, J. Math. Phys. {\bf 5},332 (1964).
\bibitem{dir31}
P. A. M. Dirac, Proc. Roy. Soc. Lond. A {\bf 133},60 (1931).
\bibitem{rks}
S. G. Rajeev, S. Kalayana Rama and S. Sen, J. Math. Phys. {\bf 35},
2259 (1994).
\bibitem{fad76}
L. Faddeev, in {\em Methods in Field Theory}, Proc. Les Houches Summer
School Vol. XXVIII, R. Balian and J. Zinn-Justin eds.
(North-Holland, Amsterdam, 1976).
\bibitem{fun94}
K. Funahashi, T. Kashiwa, S. Sakoda, K. Fujii, J. Math. Phys., in press.
\bibitem{gs}
V. Guillemin and S. Sternberg, {\em Symplectic Techniques in Physics}
(Cambridge University Press, London, 1990).
\end{thebibliography}
\end{document}